\documentclass[a4paper]{jpconf}

\usepackage[utf8]{inputenc}
\usepackage[english]{babel}

\usepackage{lmodern}
\usepackage[only,llbracket,rrbracket,llparenthesis,rrparenthesis,boxslash]{stmaryrd} % additional math symbols
\usepackage{iopams}

\usepackage{hyperref} % load this before geometry, autodetects pdftex/xetex

\newcommand{\im}{\mathrm{i}}

\newcommand{\R}{\mathbb{R}}

\newcommand{\tens}{\otimes}
\newcommand{\id}{\mathrm{id}}
\newcommand{\xd}{\mathrm{d}}

\newcommand{\cH}{\mathcal{H}}
\newcommand{\pr}{\mathcal{P}}
\newcommand{\prp}{\mathcal{P}^+}
\newcommand{\lv}{\llbracket}
\newcommand{\rv}{\rrbracket}
\newcommand{\lb}{\llparenthesis}
\newcommand{\rb}{\rrparenthesis}
\newcommand{\cb}{\mathcal{B}}
\newcommand{\cbp}{\mathcal{B}^+}
\newcommand{\cbn}{\mathcal{B}^{\mathbf{n}}}
\newcommand{\comp}{\diamond}

\newcommand{\ou}{\mathbf{e}}
\newcommand{\fn}{\mathcal{C}}

\begin{document}

\title{Born rule and Schrödinger equation\\ from first principles}
\author{Robert Oeckl}
\address{Centro de Ciencias Matemáticas,\\
  Universidad Nacional Autónoma de México,\\
  C.P.~58190, Morelia, Michoacán, Mexico}
\ead{robert@matmor.unam.mx}

\begin{abstract}
A concise review of the derivation of the Born rule and Schrödinger equation from first principles is provided. The starting point is a formalization of fundamental notions of measurement and composition, leading to a general framework for physical theories known as the positive formalism. Consecutively adding notions of spacetime, locality, absolute time and causality recovers the well established convex operational framework for classical and quantum theory. Requiring the partially ordered vector space of states to be an anti-lattice one obtains quantum theory in its standard formulation. This includes Born rule and Schrödinger equation.

\end{abstract}

\section{Introduction}

While many of the mathematical ingredients to describe classical physics are accepted as natural, this is much less so in the case of quantum physics. In particular, a complex Hilbert space, the notion of operators as observables, the Schrödinger equation and the Born rule might initially appear as lacking motivation from first principles. Indeed, many of these structures were originally introduced in a somewhat ad hoc fashion. However, as our understanding of the foundations of physics and quantum theory in particular has advanced, much of the mystery surrounding these structures has disappeared. I think it is fair to say that any mystery that remains today may be reduced to the appearance of the complex Hilbert space alone, with all other structures following naturally. While outside the scope of this note, I should add that for this last mysterious ingredient there has been a concerted effort, particularly in the last 20 years or so to explain it. These attempts, known as \emph{reconstructions of quantum theory} succeed in predicting the appearance of a complex Hilbert space by introducing some additional assumption that might appear more or less plausible.

The purpose of the present note is to review very briefly how the mathematical structures and interpretational rules of quantum theory arise naturally. I hope that especially non-experts in the foundations of quantum theory might find this account useful amid sometimes confusing and contradictory claims in the literature.
The starting point is a minimalist account of notions of measurement and composition, leading to a framework for encoding theories of physics, known as the \emph{positive formalism} \cite{Oe:posfound}, in its \emph{abstract} version. Adding a notion of spacetime and implementing the principle of locality leads to the \emph{local} version of the positive formalism (which is the starting point in \cite{Oe:posfound}). Rigidifying a \emph{metric background time} and implementing the principle of \emph{causality} leads to the \emph{convex operational framework}. The latter has been used at least since 1970 to describe classical and quantum dynamical systems in a unified fashion \cite{DaLe:opapquantprob}. The remaining steps as presented (starting with Section~\ref{sec:convop}) to recover either a rudimentary formulation of classical statistical physics or the standard formulation of quantum theory are well known. However, I have not found a presentation elsewhere in the literature that is equally condensed. Many technical details are simplified or omitted, as are most references. The interested reader is directed instead to consider \cite{Oe:posfound} and the references therein.

\section{The abstract positive formalism}

Let us start by subsuming notions of experiment, measurement, observation, preparation and intervention into a single notion, called \emph{probe}. (In the recent quantum foundations literature the word \emph{process} is often used with a similar meaning.)
A probe is a fine-grained notion in the sense that it allows, put does not require, to distinguish different measurement outcomes. For example, suppose a measurement has two distinct possible outcomes \textsf{A} and \textsf{B}. Then, there is a probe encoding each of these, call them $P(\textsf{A})$ and $P(\textsf{B})$. But there is also a probe $P(*)$ encoding the experiment without outcome, i.e., where no knowledge of a specific outcome is taken into account. In general, an experiment may involve a number of output channels, each of which may or may not be observed in a given run. There is then a probe encoding each of these different possibilities in addition to each of the possible outcomes of the observed output channels. This leads to a \emph{hierarchy of generality} in the space of probes that gives rise to a binary \emph{relation} of relative generality. It is not difficult to see that the properties of this binary relation (reflexive, anti-symmetric and transitive) are precisely those of a \emph{partial order}. Hence we shall say that a probe $P$ that is more special than a probe $Q$, i.e., implies $Q$, is less or equal to $Q$, $P\le Q$. In the simple example with two outcomes we have $P(\textsf{A})\le P(*)$ and $P(\textsf{B})\le P(*)$.

It is not sufficient to consider single experiments or their outcomes in isolation. Meaningful predictions arise in terms of correlations between different observations and interventions. To capture this, let us introduce a notion of \emph{composition} of probes, denoted $P\comp Q$, for probes $P$ and $Q$. For example, if the component probes encode certain outcomes of different experiments, then the composite probe encodes the joint occurrence of all of these outcomes. Not all probes can be meaningfully composed. For example, a probe involving a certain apparatus at a certain place and time cannot be composed with a probe involving a different (or even a copy of the same) apparatus at the same place and time.

In general it is not practical to consider explicitly for a given experiment all observations or interventions taking place that are in some way correlated to this experiment. Instead, let us introduce a notion of \emph{boundary condition} that parametrizes all potential correlations between the experiment and the rest of the world. So, associated to any given probe is a set of boundary conditions. This set may be different for different probes, but it is the same for probes that encode different aspects (such as different outcomes) of the same run of the same experiment. Let us say that probes that share the same set of boundary conditions are of the same \emph{type}. In particular, there are partial order relations only between probes of the same type. On the other hand, probes of the same type cannot be composed with each other.

The most elementary prediction one might wish to make is the \emph{probability} for a certain outcome of an observation or measurement. Consider the previous example of a single experiment with two possible outcomes \textsf{A} and \textsf{B}. Say, we are interested in the probability $\Pi$ for outcome \textsf{A}. Clearly we need the probe $P(\textsf{A})$ to encode this outcome. However, and this is less obvious, we also need the probe $P(*)$ as a reference for comparison. Moreover, we need a boundary condition $b$ that parametrizes our knowledge about the influence of all the external degrees of freedom on our experiment. We may then loosely paraphrase the question we ask as follows: How probable is the observation of $P(\textsf{A})$ subject to boundary condition $b$, given that we have set up experiment $P(*)$ subject to boundary condition $b$? We translate this into a mathematical formula as follows,
\begin{equation}
  \Pi =\frac{\lv P(\textsf{A}),b\rv}{\lv P(*),b\rv} .
  \label{eq:elemprob}
\end{equation}
Here we have introduced a pairing $\lv\cdot,\cdot\rv$ between probes of a given type and corresponding boundary conditions. The pairing takes non-negative real values. These values are not in general probabilities, but can be thought of as ``relative probabilities'' or \emph{compatibilities}. So, $\lv P(*),b\rv$ quantifies the compatibility between the experimental setup and the boundary condition. $\lv P(\textsf{A}),b\rv$ quantifies this same compatibility, but where in addition the occurrence of outcome $\textsf{A}$ is imposed. Crucially, the inequality $P(\textsf{A})\le P(*)$ encoding relative generality translates precisely into a corresponding inequality of compatibilities:
\begin{equation}
  \lv P(\textsf{A}),b\rv \le \lv P(*),b\rv .
\end{equation}
This guarantees that the quotient (\ref{eq:elemprob}) is less or equal to one, as a probability must be. Note that this inequality will hold for any boundary condition $b$.

Conversely, we may use this to define the partial order relation in the space of probes of a given type. This amounts to viewing probes, via the pairing, as real valued functions on a set (the set of boundary conditions). The real vector space of real valued functions on a set naturally carries a partial order compatible with the vector space structure. It is thus called a \emph{partially ordered vector space}. There holds $f\le g$ for functions $f$ and $g$ if $f(x)\le g(x)$ for all elements $x$ of the set. We denote the partially ordered vector space spanned by all probes of a given type by $\pr$. We denote the subset of \emph{positive elements} of $\pr$ by $\prp$. We call an element \emph{positive} if it satisfies $0\le P$. This subset is in fact a \emph{convex cone}. In particular, all probes are positive since compatibilities are non-negative. Returning to the example, we also have a completeness condition that assures that probabilities add up to one. Here, for any boundary condition $b$,
\begin{equation}
  \lv P(\textsf{A}),b\rv + \lv P(\textsf{B}),b\rv = \lv P(*),b\rv .
\end{equation}
This translates to, $P(\textsf{A})+P(\textsf{B})=P(*)$.

What is more, we shall assume that all elements of $\prp$ are probes. This may be justified as follows. Firstly, we claim that performing a probabilistic mixture of experiments is also a valid experiment. Thus, probabilistic mixtures of probes are probes. Secondly, we notice a rescaling freedom for probes in expression (\ref{eq:elemprob}). (Only relative scalings are important.) Thus a positive multiple of a probe is again a probe. Consequently, linear combinations of probes with positive coefficients are probes. In this way we obtain the whole cone $\prp$. In contrast, we call general elements of $\pr$ \emph{generalized probes}.\footnote{What is called a probe here is called a \emph{primitive probe} in \cite{Oe:posfound}. A generalized probe here is simply a probe in \cite{Oe:posfound}.}
Similarly, it is legitimate to form probabilistic mixtures of boundary conditions and expression (\ref{eq:elemprob}) reveals a scaling freedom also for these. In this way, the boundary conditions form a positive cone $\cbp$ in a partially ordered vector space $\cb$ of \emph{generalized boundary conditions}. The pairing $\lv \cdot,\cdot \rv$ thus becomes a map $\prp\times\cbp\to\R^+_0$ and we extend it bilinearly to a map $\pr\times\cb\to\R$.

We take a closer look at the interplay of composition and boundary conditions. Suppose we consider the composition of a probe $P$ with a probe $Q$. We recall that the boundary conditions for the probe $P$ encode knowledge about correlations between $P$ and the rest of the world. Now $Q$ is part of this rest of the world. Thus, we should expect part of these boundary conditions to encode correlations with $Q$ and another part correlations with the rest of the world excluding $Q$. Corresponding considerations hold for $Q$. To formalize this we introduce a notion of decomposition of the spaces of boundary conditions. On the level of generalized boundary conditions which are vector spaces, this decomposition should take the form of a tensor product. This is again because we permit probabilistic combinations as well as rescalings. In the example with probes $P$ and $Q$, call the corresponding spaces of generalized boundary conditions $\cb_P$ and $\cb_Q$. We denote the decompositions by $\cb_P=\cb_{PQ}\tens\cb_P'$ and $\cb_Q=\cb_{PQ}\tens\cb_Q'$. Crucially, the space $\cb_{PQ}$ of boundary conditions encoding correlations between $P$ and $Q$ appears in both decompositions. What is more, $P$ and $Q$ are composed ``along'' $\cb_{PQ}$. The joint probe $P\comp Q$, arising from this composition has a space of boundary conditions that decomposes as $\cb=\cb_{P}'\tens\cb_{Q}'$.

We allow a special \emph{transparent} probe that does not correspond to any actual intervention, measurement or observation, but just ``lets signals pass through'' from any space $\cb$ of boundary conditions to a copy of this same space. Using this we can derive an inner product $\lb\cdot,\cdot\rb:\cb\times\cb\to\R$ that is positive-definite and non-negative on pairs of positive elements (proper boundary conditions). This in turn leads to the \emph{composition rule} of probes as follows. We return to the example of probes $P$ and $Q$ with corresponding spaces of boundary conditions $\cb_P=\cb_{PQ}\tens\cb_P'$ and $\cb_Q=\cb_{PQ}\tens\cb_Q'$. Let $b\in \cb_P'$ and $c\in\cb_Q'$. Let $\{\xi_k\}_{k\in I}$ be an orthonormal basis of $\cb_{PQ}$. Then,
\begin{equation}
  \lv P\comp Q, b\tens c\rv=\sum_{k\in I} \lv P, b\tens\xi_k\rv \lv Q, c\tens\xi_k\rv .
  \label{eq:pcomp}
\end{equation}

We arrive at an \emph{axiomatic} framework for physical theories, called the \emph{abstract positive formalism}. Its main mathematical ingredients, called probes and boundary conditions, are \emph{partially ordered vector spaces}. Crucially this framework includes rules for the \emph{prediction of physically measurable quantities}, the simplest of those taking the form of formula (\ref{eq:elemprob}). A \emph{physical theory} now consists of a model satisfying the axioms, i.e., an explicit specification of spaces of probes, boundary conditions, types etc.\ along with a correspondence to physical phenomena for some of these objects.

\section{Spacetime and locality}

The framework as considered so far is meant to be general and minimalistic, involving as few assumptions about our concrete experience of the physical world as possible. However, adding suitable structure that codifies (perhaps a simplified version of) this experience can greatly improve the practical usefulness and predictive power of the framework when applicable. We proceed here to integrate a notion of \emph{spacetime} into the framework, which allows to implement the powerful principle of \emph{locality}.

We suppose that measurements can be localized in \emph{spacetime regions}. In fact, we identify the \emph{types} of probes with spacetime regions. In this way we assign to each spacetime region $M$ its corresponding space $\pr_M$ of (generalized) probes. We permit the composition of probes only if the underlying spacetime regions have no intersection, except in their boundaries. That is, the composition of probes corresponds to a composition of the underlying spacetime regions in terms of a \emph{gluing} operation. Similarly, we associate the space of (generalized) boundary conditions for probes in $M$ to the boundary hypersurface $\partial M$ of the region $M$, with notation $B_{\partial M}$. More generally, we allow the association of a space $\cb_{\Sigma}$ of boundary conditions to any hypersurface $\Sigma$. Moreover, we bring into correspondence decompositions of spaces of boundary conditions with decompositions of the underlying hypersurfaces. Thus, a decomposition of a hypersurface $\Sigma$ is its presentation as a union $\Sigma=\Sigma_1\cup\cdots\cup \Sigma_n$ of hypersurfaces such that the components may intersect only in their boundaries. To this corresponds then the tensor product decomposition of associated spaces of boundary conditions, $\cb_{\Sigma}=\cb_{\Sigma_1}\tens\cdots\tens\cb_{\Sigma_n}$. The composition of probes in adjacent spacetime regions involves the decomposition of the boundary hypersurface of each involved region into pieces. Each piece arises as the intersections with the boundaries of one of the other regions. Returning to the example of composing probes $P$ and $Q$, these are now associated to underlying spacetime regions $M$ and $N$. The boundaries $\partial M$ and $\partial N$ intersect in some hypersurface $\Sigma_{PQ}$. This leads to decompositions $\partial M=\Sigma_{PQ}\cup \Sigma_{P}$ and $\partial N=\Sigma_{PQ}\cup \Sigma_{Q}$, where $\Sigma_P$ and $\Sigma_Q$ are the non-intersecting pieces of the boundaries. We recover the previously considered decomposition the spaces of boundary conditions as follows, $\cb_P=\cb_{\partial M}$, $\cb_Q=\cb_{\partial N}$, $\cb_{PQ}=\cb_{\Sigma_{PQ}}$, $\cb_P'=\cb_{\Sigma_P}$, $\cb_Q'=\cb_{\Sigma_Q}$.

Crucially, two probes only share a space of boundary conditions if the underlying regions are adjacent, that is if parts of their boundaries intersect. This implements the powerful principle of \emph{locality} which dramatically reduces the possible direct connections between probes in a composition. The underlying physical idea is that interaction is exclusively mediated by signals traversing spacetime and thereby necessarily crossing interfacing hypersurfaces.
For each spacetime region there is a special kind of probe that corresponds to not making any intervention, measurement or observation in this region. We call this the \emph{nul probe}. Clearly, the composition of nul probes is a nul probe.
We arrive at the \emph{local positive formalism}, also called \emph{spacetime positive formalism} (and previously simply referred to as the \emph{positive formalism}). It is at this level that we may make contact with classical field theory or quantum field theory \cite{Oe:posfound}. However, as we are interested in recovering the standard formulation of quantum theory we proceed to specialize to a non-relativistic picture of spacetime.

\section{Time, evolution and causality}
\label{sec:tevolcs}

We drastically simplify the notion of spacetime to a linear notion of time, measured in terms of real numbers. Spacetime regions are reduced to time-intervals $[t_1,t_2]$ and finite unions of these. Hypersurfaces are reduced to instants of time (and finite unions thereof). In particular, we have a space of boundary conditions $\cb_t$ for each time $t\in\R$. Any other space of boundary conditions arises as a tensor product of these. We also call $\cb_t$ the \emph{state space} at time $t$ and its elements \emph{(generalized) states}. Intuitively speaking, a state parametrizes knowledge about the past relevant for making predictions about the future. But it also parametrizes information about the future relevant for making predictions about the past. What is more, by construction, any possibly relevant information about the past (or future) may be encoded in this way.

Probes are associated to time intervals and admit a pairing with boundary conditions that arise from tensor products of initial and final states. However, it turns out to be convenient to view probes instead as maps from the initial to the final state space. This is equivalent via the inner product. Consider a time interval $[t_1,t_2]$ and a corresponding probe $P\in\prp_{[t_1,t_2]}$. Set $\tilde{P}:\cb_{t_1}\to\cb_{t_2}$ via,
\begin{equation}
  \lv P, b_1\tens b_2\rv_{[t_1,t_2]}=\lb b_2, \tilde{P}\, b_1\rb_{t_2} \quad\forall b_1\in\cb_{t_1}, \forall b_2\in\cb_{t_2} .
\end{equation}
$\tilde{P}$ is uniquely determined by $P$ and vice versa. Note crucially that the positivity of $P$ implies that $\tilde{P}$ maps positive elements to positive elements, i.e., restricts to a map $\cbp_{t_1}\to\cbp_{t_2}$. (The converse is not necessarily true, see \cite{Oe:posfound} for more details.)
From this point onwards we exclusively use the presentation of probes as maps between state spaces and refer to them as \emph{operations}. (We also drop the tilde in the notation, i.e., write $P$ to mean $\tilde{P}$.) This has the advantage that the somewhat cumbersome composition rule for probes (\ref{eq:pcomp}) translates into a simple composition of maps.
Recall the special nul probe (or operation) for any time interval $[t_1,t_2]$ which represents the absence of any intervention, measurement or observation during that time. We denote this by $T_{[t_1,t_2]}:\cb_{t_1}\to\cb_{t_2}$ and call it the \emph{time evolution} map. Time evolution maps compose as $T_{[t_1,t_3]}=T_{[t_2,t_3]}\circ T_{[t_1,t_2]}$ for $t_1\le t_2\le t_3$ because nul probes compose to nul probes.

We introduce a further simplification at his point, namely we suppose that all the state spaces $\cb_t$ are isomorphic and can be canonically identified with a single state space $\cb$ via a time translation symmetry. Operations are then positive operators on $\cb$, i.e., linear maps $\cb\to\cb$ that preserve positivity. Further, we refer from here onwards to a generic space of operations $\prp\subseteq\pr$ without necessarily specifying an initial and final time for each operation. Let us consider the form the probability formula (\ref{eq:elemprob}) takes with the newly introduced notation. Thus, we consider a measurement with two possible outcomes \textsf{A} and \textsf{B}. The operations $P(\textsf{A})$ and $P(*)$ represent the measurement with specific outcome \textsf{A} and unspecified outcome respectively. We need to specify an initial state $b_1\in\cbp$ and a final state $b_2\in\cbp$ to obtain a probability $\Pi$ for the occurrence of outcome \textsf{A},
\begin{equation}
  \Pi=\frac{\lb b_2, P(\textsf{A})\, b_1\rb}{\lb b_2, P(*)\, b_1\rb} .
  \label{eq:tevprob}
\end{equation}
Here, $b_1$ encodes knowledge about the past of the measurement and $b_2$ knowledge about its future. The choice of $b_1$ is usually referred to as a \emph{preparation}. A choice of the state $b_2$ on the other hand is usually called \emph{post-selection}.

But what if we do not have any knowledge about the future (or past)? To codify this situation we distinguish a special state $\ou\in\cbp$. Recall how the partial order structure of $\cbp$ corresponds to a hierarchy of generality. What we want is a ``maximally general'' state, a \emph{state of maximal uncertainty}. In terms of the partial order this should mean something like $b\le\ou$ for any other state $b$. Of course this cannot be literally true, because we have rescaling freedom. But taking this into account we can demand the inequality after suitable rescaling. That is, we demand that for any state $b\in\cbp$ there exist $\lambda>0$ such that $b\le \lambda\ou$. Such an element of a partially ordered vector space is called an \emph{order unit}. Our distinguished state $\ou$ is thus a particular order unit (in general there are many). We are now in a position to talk about a measurement without post-selection. This is achieved by setting the final state to be the state of maximal uncertainty, $b_2=\ou$ in formula (\ref{eq:tevprob}). This is also referred to as \emph{discarding} the system that has been measured. Note that we could equally contemplate using the state of maximal uncertainty as an initial state or even introducing this concept already at the level of the local or abstract positive formalism, but this is outside of the scope of this short piece.

The concept of the state of maximal uncertainty gives us the opportunity to hard-code another well established empirical principle into the formalism, namely \emph{causality}. By this we mean that a future choice of measurement (but not outcome!) does not influence outcomes in the past. To this end we distinguish \emph{non-selective} from \emph{selective} operations. Thus, consider again a measurement with two possible outcomes \textsf{A} and \textsf{B} and corresponding operations $P(\textsf{A})$, $P(\textsf{B})$, $P(*)$. Then, the operation $P(*)$ is non-selective as it only represents performing the measurement without knowledge of the outcome, while the operations $P(\textsf{A})$ and $P(\textsf{B})$ are selective as they select outcome \textsf{A} or \textsf{B}.

Consider now the consecutive execution of two measurements of this type, first $P$, then $Q$. Given a fixed initial state $b\in\cbp$, the probability $\Pi$ to measure outcome \textsf{A} via $P$ and then discarding should be the same as when performing $Q$ (without getting to know its outcome) after $P$ and only then discarding. That is,
\begin{equation}
  \Pi=\frac{\lb \ou, Q(*) P(\textsf{A})\, b\rb}{\lb \ou, Q(*) P(*)\, b\rb}=\frac{\lb \ou, P(\textsf{A})\, b\rb}{\lb \ou, P(*)\, b\rb} .
\end{equation}
For this to hold with arbitrary $b$ and $P$ requires the \emph{normalization condition} for non-selective operations,
\begin{equation}
  \lb\ou, Q(*) c\rb=\lb\ou, c\rb,\quad\forall c\in\cbp .
  \label{eq:nselnorm}
\end{equation}
For selective probes this implies a corresponding inequality, e.g., $Q(\textsf{A})\le Q(*)$ implies, $\lb\ou,Q(\textsf{A}) c\rb \le \lb\ou,Q(*)c\rb=\lb\ou,c\rb$. From this point onwards we implement the causality principle by requiring non-selective operations to satisfy the normalization condition (\ref{eq:nselnorm}). Note that this condition is \emph{time asymmetric}, in contrast to all other elements of the formalism considered so far.

Non-degeneracy and positivity of the inner product imply that the inner product between the state of maximal uncertainty $\ou$ and an arbitrary state $b\in\cbp$ that is non-zero, $b\neq 0$, is strictly positive $\lb \ou, b\rb> 0$. We can take advantage of this to introduce a \emph{normalization condition} for states. We say that a state $b\in\cbp$ is \emph{normalized} if $\lb\ou,b\rb=1$. Thus, any non-zero state can be normalized by multiplying it with a positive scalar. We denote the subset of normalized states by $\cbn\subseteq\cbp\subseteq\cb$. Using normalized states when the normalization condition for non-selective operations is enforced leads to simplifications of some formulas. Most notably, the probability for the outcome of a sequence of measurements takes the form of a linear expression. However, this only works if the system is discarded at the end and no conditioning on outcomes is considered. For a single measurement $P$ the probability $\Pi$ of outcome \textsf{A} with initial state $b\in\cbn$ takes the simple form,
\begin{equation}
  \Pi=\lb \ou,P(\textsf{A})\, b\rb .
  \label{eq:simplprob}
\end{equation}
This arises from formula (\ref{eq:tevprob}) with $b_2=\ou$, $b_1=b\in\cbn$ because the denominator takes the value $1$, as $P(*)$ satisfies (\ref{eq:nselnorm}).

If we perform consecutive measurements we can ask how we should ``update'' a state to codify the outcome of a previous measurement. Starting with a state $b\in\cbp$ and performing a measurement $P$ with outcome $\textsf{A}$, the probability for an outcome $\textsf{B}$ in a subsequent measurement $Q$ is,
\begin{equation}
  \Pi=\frac{\lb\ou, Q(\textsf{B}) P(\textsf{A}) b\rb}{\lb\ou, Q(*) P(\textsf{A}) b\rb}= \frac{\lb\ou, Q(\textsf{B}) c\rb}{\lb\ou, Q(*) c\rb} .
\end{equation}
Here we have set $c=P(\textsf{A}) b$. This can be seen as the updated state. However, if we want to take advantage of simplified formulas such as (\ref{eq:simplprob}) we have to work with normalized state. This yields the \emph{state update rule},
\begin{equation}
  b\mapsto \frac{P(\textsf{A})\, b}{\lb\ou,P(\textsf{A})\, b\rb} .
  \label{eq:supdate}
\end{equation}

\section{The convex operational framework}
\label{sec:convop}

The framework we have arrived at is the well known \emph{convex operational framework}, designed to describe classical and quantum theory in a unified way \cite{DaLe:opapquantprob}. The main ingredients of this framework are thus a partially ordered vector space $\cb$ called (generalized) \emph{state space} and a partially ordered vector space $\pr$ of (generalized) \emph{operations}. The elements of $\pr$ are linear operators on $\cb$ with the subset $\prp$ of (proper) operations mapping states to states, i.e., $\cbp$ to itself. Moreover, there is a special \emph{state of maximal uncertainty} $\ou\in\cbp$ which is an order unit. We have a subset $\cbn\subseteq\cbp$ of \emph{normalized states} $b$ that satisfy $\lb\ou,b\rb=1$. Operations are non-selective or selective with the former satisfying the normalization condition (\ref{eq:nselnorm}). Probabilities for outcomes of single measurements with out post-selection take the form (\ref{eq:simplprob}) and come with the state update rule (\ref{eq:supdate}).

The evolution of states in the absence of any measurement or intervention is encoded in the time evolution maps $T_{[t_1,t_2]}\in\prp$, associated to time intervals $[t_1,t_2]$. These compose in the obvious way, as already remarked. We shall assume now in addition that the time evolution maps $T_{[t_1,t_2]}$ preserve the structure of the state space completely. This is reasonable if we are considering a complete description of physics in the sense that we are not neglecting any interaction with an ``exterior'' system. Thus, $T_{[t_1,t_2]}:\cb\to\cb$ is an isomorphism of vector spaces and reduces to a bijection of positive cones $\cbp\to\cbp$. Moreover, it preserves the inner product $\lb\cdot,\cdot\rb$ and the state of maximal uncertainty, $T_{[t_1,t_2]}\ou=\ou$ which implies that it also preserves the subset of normalized states $\cbn$. We make now another simplifying assumption about the physics, namely that time evolution is itself time translation invariant. That is, $T_{[t_1,t_1+\Delta]}=T_{[t_2,t_2+\Delta]}$ for any $t_1,t_2,\Delta$. Consequently, we simply write $T_\Delta$ instead of $T_{[t_1,t_1+\Delta]}$. In this way we obtain a one-parameter group of isomorphisms of $\cb$, i.e., satisfying $T_{\Delta_1} T_{\Delta_2}=T_{\Delta_1+\Delta_2}$ and $T_0=\id$.

\section{Classical theory}

After this lightning review of the convex operational framework as derived from first principles via the positive formalism we come to the question of how we recover the details of the standard formulation of quantum theory from it. The answer is that, of course, we cannot do so yet as we have not introduced any principle that would exclude, say, classical physics. Indeed, not all partially ordered vector spaces are alike, even with the additional structures they carry in our case. (This is in notable contrast to, say, infinite-dimensional separable complex Hilbert spaces.) It is thus not surprising that radically different models of physics can be captured with this framework.

We start with a particular type of partially ordered vector space as state space, a \emph{lattice}. Roughly speaking this means that for any two positive elements $b,c\in\cbp$ there is a positive element that represent the minimum $\min(b,c)$ and one that represents the maximum $\max(b,c)$ of the two elements. This turns out to imply that $\cb$ is the vector space $\fn(L)$ of functions over a set $L$.
This set $L$ is nothing but the \emph{phase space} of a \emph{classical theory}. $\cbp$ is the subset of positive functions, i.e., the functions that have everywhere non-negative values. The physical interpretation of the states is that of \emph{statistical distributions} on phase space. Without explaining details let us quickly go over the other structures. There is also a \emph{measure} $\mu$ on $L$ and the inner product on $\cb$ is the corresponding $L^2$-inner product,
\begin{equation}
  \lb b,c\rb=\int b(x) c(x) \xd\mu(x) .
\end{equation}
The state $\ou$ of maximal uncertainty is the constant function of unit value. (We assume that all functions are bounded.) The normalization condition $\lb\ou,b\rb$ for a state $b\in\cbp$ means that it has unit integral. The operations are positive maps $\cb\to\cb$ which fall into classes with different physical interpretations. An interesting class is given by the particular \emph{observables} that are characteristic functions $\chi_S$ for subsets $S\subseteq L$ of the phase space. This yields a selective operation as the map $b\mapsto \chi_S\cdot b$. The corresponding non-selective operation is the identity operator. This map encodes a measurement to determine whether the system is in the subset $S$. The probability for an affirmative answer is given by formula (\ref{eq:simplprob}) and the corresponding update of the statistical distribution is the state update rule (\ref{eq:supdate}).
Another interesting class of operations contains those that preserve the structure of $\cb$ completely. It turns out that these are necessarily induced by measure preserving bijections $v:L\to L$ and take the form, $b\mapsto b\circ v^{-1}$. The time evolution maps are of this type. In particular, they arise thus from a one-parameter group of measure preserving bijections $v_{\Delta}$ in phase space. This flow in phase space can be described infinitesimally by a vector field. Usually there is a symplectic structure on phase space so that this is the Hamiltonian vector field of the Hamiltonian function.

\section{Quantum theory}

Another type of partially ordered vector space arises as the space of \emph{self-adjoint operators} on a complex Hilbert space $\cH$ with inner product denoted by $\langle\cdot,\cdot\rangle$. One way to characterize this type of space is as an \emph{anti-lattice}. That is, for any two positive elements a minimum or maximum exists only in the trivial case when the elements are linearly dependent. Thus, the generalized state space $\cb$ consists of self-adjoint operators on $\cH$. $\cbp$ is the subset of \emph{positive operators}, i.e., operators $b$ such that $\langle b\psi,\psi\rangle\ge 0$ for all $\psi\in\cH$. The inner product on $\cb$ is then necessarily the Hilbert-Schmidt inner product,
\begin{equation}
  \lb b, c\rb = \tr(b c) ,
\end{equation}
and the state of maximal uncertainty is $\ou=\id$, the identity operator (both up to suitable equivalence). The state normalization condition acquires the simple form $\lb\ou,b\rb=\tr(b)=1$ for $b\in\cbp$. In particular, the subset $\cbn\subseteq \cbp$ is thus precisely the set of positive operators on $\cH$ of unit trace. The set $\prp$ of operations is the set of \emph{completely positive maps} $\cb\to\cb$. (In the language of Section~\ref{sec:tevolcs}: A generalized probe is positive if and only if the associated probe map is completely positive.) In the present context they are called \emph{quantum operations}. Such an operation $Q$ can be written in terms of Kraus operators $\{K_i\}_{i\in I}$ as, $Q b = \sum_i K_i b K_i^\dagger$. The normalization condition for non-selective operations (\ref{eq:nselnorm}), is easily seen to be equivalent to $\tr(\sum_i K_i^\dagger K_i)=1$.
The special case of a measurement given by an \emph{observable} $A$, i.e., a self-adjoint operator on $\cH$ is as follows. Let $\sum_i a_i P_i$ be the spectral decomposition of $A$ in terms of orthogonal projectors $P_i$. (For simplicity we suppose a finite sum.) $a_i$ are the distinct real numbers labeling the distinct measurement outcomes. The non-selective operation $Q(*)$ associated with this measurement is the map $b\mapsto \sum_i P_i b P_i$ while the selective operation $Q(k)$ selecting outcome $k$ is, $b\mapsto P_k b P_k$. With this the probability (\ref{eq:simplprob}) for outcome $k$ in state $b\in\cbn$ becomes,
\begin{equation}
  \Pi=\lb \ou, Q(k) b\rb=\tr(P_k b) .
\end{equation}
If $b$ is a pure state, i.e., can be written as $b=|\psi\rangle\langle\psi|$ for $\psi\in\cH$ in the usual bra-ket notation, then this is also the same as,
\begin{equation}
  \Pi=\langle\psi, P_k\psi\rangle .
\end{equation}
If $P_k$ is a one-dimensional projector, i.e., of the form $P_k=|\eta_k\rangle\langle\eta_k|$, then this in turn can be written as,
\begin{equation}
  \Pi=|\langle\eta_k,\psi\rangle|^2 .
\end{equation}
This is precisely the Born rule.
The state transformation rule (\ref{eq:supdate}) in case of outcome $k$ is,
\begin{equation}
  b\mapsto\frac{P_k b P_k}{\tr(P_k b)},\qquad\mathrm{or}\qquad \psi\mapsto \frac{P_k\psi}{\sqrt{\langle\psi, P_k\psi\rangle}},
\end{equation}
in terms of pure states if applicable. The latter expression is justified by reproducing the former and involves a convenient, but irrelevant choice of phase. In particular, we recover the Lüders rule.

We proceed to consider operations that preserve the structure of the state space, including its inner product and state of maximal uncertainty. These have necessarily the form $b\mapsto U b U^\dagger$, where $U$ is a unitary operator on $\cH$. In particular, for the one-parameter group of time-evolution maps $T_{\Delta}$ this means that (fixing phases as necessary) there is a corresponding one-parameter group of unitary maps $U_{\Delta}$ such that $T_{\Delta} b= U_{\Delta} b U_{\Delta}^\dagger$. This satisfies $U_{\Delta_1} U_{\Delta_2}=U_{\Delta_1+\Delta_2}$ and $U_0=\id$. By Stone's theorem there exists then a self-adjoint operator $H$ on $\cH$ such that $U_{\Delta}=\exp(-\im \Delta H)$, called the \emph{Hamiltonian operator}. Adopting the notation $b(t)=T_t b=U_t b U_t^{-1}$ we can then write the time evolution of the state $b(t)$ as a differential equation,
\begin{equation}
  \frac{\partial}{\partial t} b(t)=-\im [H,b(t)] ,
\end{equation}
where $[\cdot,\cdot]$ denotes the commutator. This is precisely the \emph{quantum Liouville equation}, i.e., the special case of the \emph{Lindblad equation} for closed systems. For a pure state $b(t)=|\psi(t)\rangle\langle\psi(t)|$ we can set $|\psi(t)\rangle=U_t |\psi(0)\rangle$. In differential form,
\begin{equation}
  \frac{\partial}{\partial t} |\psi(t)\rangle=-\im H |\psi(t)\rangle .
\end{equation}
This is the Schrödinger equation.

\ack This work was partially supported by CONACYT project grant 259258 and UNAM-PAPIIT project grant IA-106418.

\section*{References}

\bibliographystyle{iopart-num} % bibliography
\bibliography{stdrefsb}
\end{document}